\documentclass[a4paper]{jpconf}

\usepackage{graphicx}
\usepackage{amsmath}
\usepackage{bm}
\usepackage{mathrsfs}
\usepackage{color}
\usepackage{slashed}
\usepackage{dcolumn}% Align table columns on decimal point

\newcommand{\bald}[1]{{\bf #1}}

\newcommand{\cur}[1]{\mathscr{ #1}}
\newcommand{\eqf}[1]{\begin{equation}\begin{split}#1\end{split}\end{equation}}

\begin{document}

\title{A derivation of the source term induced by a fast parton from the quark energy-momentum tensor}

\author{R. B. Neufeld}

\address{Los Alamos National Laboratory, Theoretical Division, MS B238, Los Alamos, NM 87545, U.S.A.}

\ead{neufeld@lanl.gov}

\begin{abstract}
The distribution of energy and momentum deposited by a fast parton in a medium of thermalized quarks, or the source term, is evaluated in perturbative thermal field theory.  The calculation is performed by directly evaluating the thermal expectation value of the quark energy-momentum tensor.  The fast parton is coupled to the medium by adding an interaction term to the Lagrangian.  I show that this approach is very general and can be modified to consider more realistic modeling of fast parton propagation, such as a fast parton created in an initial hard interaction or the evolution of a parton shower due to medium induced radiation.  For the scenario considered here, it is found that local excitations fall sharply as a function of the energy of the fast parton.  These local excitations couple directly to the sound mode in hydrodynamics and are important for generating an observable shockwave structure.  This may have implications for the trigger $p_T$ dependence of measurements of azimuthal dihadron particle correlations in heavy-ion collisions.  In particular, one would be less likely to observe a conical emission pattern for increasing trigger $p_T$.  
\end{abstract}

\section{Introduction}

The observation of the quark gluon plasma (QGP) created in heavy-ion collisions \cite{qgpform} is one of the most exciting discoveries of the past decade.  Among the many interesting experimental results from the heavy-ion programs at the Relativistic Heavy Ion Collider (RHIC) and the Large Hadron Collider (LHC), the suppression in the production rates of energetic, or hard, leading particles and jets relative to a naive superposition of nucleon-nucleon collisions is one of the most striking \cite{Adcox:2001jp,LHC,LHCII}.  These phenomena are often collectively referred to as ``jet quenching'' and are reflective of the interaction of hard partons with the QGP medium which they traverse.  

Another striking result from heavy-ion experiments is that the QGP formed at currently achievable center-of-mass energies may behave as a nearly ideal fluid \cite{Song:2007fn}.  In particular, the low energy, or soft particles created in heavy-ion collisions show a strong flow profile consistent with nearly ideal fluidity.  The two results cited above lead to the natural question of how the QGP responds to a propagating fast (hard) parton.  More generally stated, one would like to understand how the hard and soft physics of the QGP join together.

In this proceedings I will attempt to make some progress in that direction.  I will present a new approach for coupling the hard physics of fast parton propagation to the underlying  thermal medium.  The idea is to couple the fast parton to the QGP via an interaction term in the Lagrangian.  The dynamics of the medium in the presence of the fast parton can then be directly obtained by taking the thermal expectation value of the energy-momentum tensor (EMT), denoted $T^{\mu\nu}$.  Although I will here consider an asymptotically propagating fast parton, it should be stressed that the approach used in this proceedings can be easily extended to consider more realistic scenarios.  For instance, one can modify the interaction term to consider back to back fast partons created in an initial hard interaction, which is more relevant to phenomenology.  Or, equally as relevant, one can incorporate \cite{Neufeld:2011yh} the medium induced evolution of a primary fast parton into a full parton shower through radiative processes \cite{jet1}.

Ideally one would like to evaluate components of the EMT directly, but for the reasons discussed in \cite{Neufeld:2010xi} the focus here will instead be on the divergence of the EMT, $\partial_\mu T^{\mu\nu}$.  I will refer to this quantity as the "source term", in keeping with common terminology.  The source term carries information about the rate of energy transfer to the medium and acts as a source (hence the name "source term") for the evolution of the underlying medium in the presence of a fast parton.  By using an effective theory such as hydrodynamics, one can evaluate the medium response to a fast parton from the source term.  A rigorously derived source term thus provides guidance for phenomenological studies of shockwave formation in the QGP \cite{machphenom}.

This dual utility of the source term makes it a powerful quantity, particularly when applied to the realistic scenarios discussed above.  For instance, the amount and distribution of energy a parton shower loses to the underlying medium can be obtained from the source term.  This is useful in the description of jet observables \cite{jetty}, which are a more powerful probe of the QGP than leading particle observables.  Additionally, such information may be useful in the experimental/jet medium background separation in heavy-ion experiments \cite{LHC,LHCIII}. 

\section{Formalism and Results}
\label{form}

The approach here is to begin with the EMT of the underlying medium.  For simplicity I will consider a medium of massless quarks/antiquarks, leaving the inclusion of medium gluons for a forthcoming study.  Thus, apart from the coupling strengths, the results will be the same as for a QED plasma.  The EMT is given by \cite{qedemt}
\eqf{\label{qemt}
T^{\mu\nu} = \frac{i}{4}\bar{\psi}\left(\gamma^\mu\,\overset{\text{\tiny$\leftrightarrow$}}{D^\nu} + \gamma^\nu\,\overset{\text{\tiny$\leftrightarrow$}}{D^\mu}\right)\psi - g^{\mu\nu}\cur{L},
}
where
\eqf{
\cur{L} = \frac{i}{2}\bar{\psi}\,\overset{\text{\tiny$\leftrightarrow$}}{\slashed{D}}\,\psi \text{,     }D^\mu = \partial^\mu - i g\,A_a^\mu\,t^a
}
and
\eqf{
\bar{\psi}\,\gamma^\mu\,\overset{\text{\tiny$\leftrightarrow$}}{D^\nu} \, \psi = \bar{\psi}\,\gamma^\mu\,\overset{\text{\tiny$\rightarrow$}}{D^\nu} \, \psi - \bar{\psi}\,\gamma^\mu\,\overset{\text{\tiny$\leftarrow$}}{D^{*\nu}} \, \psi.
}
In the above equations, $g$ is the strong coupling, $t^a$ are the $SU(3)$ generators in the fundamental representation, and conventional slashed notation is used, $\slashed{A} = \gamma_\mu A^\mu$, etc.  A summation over color, spin, and the active number of quark flavors is implied in the EMT.

\begin{figure*}
\centerline{
\includegraphics[width = 0.25\linewidth]{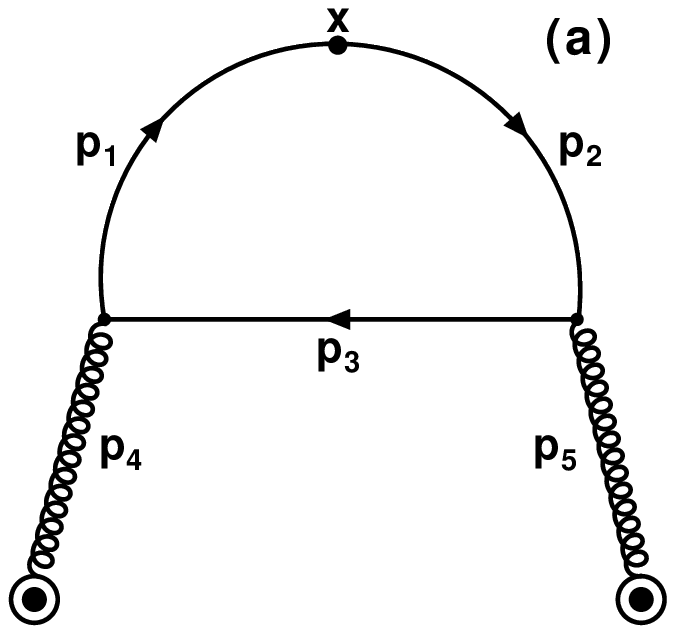}\hskip0.15\linewidth
\includegraphics[width = 0.26\linewidth]{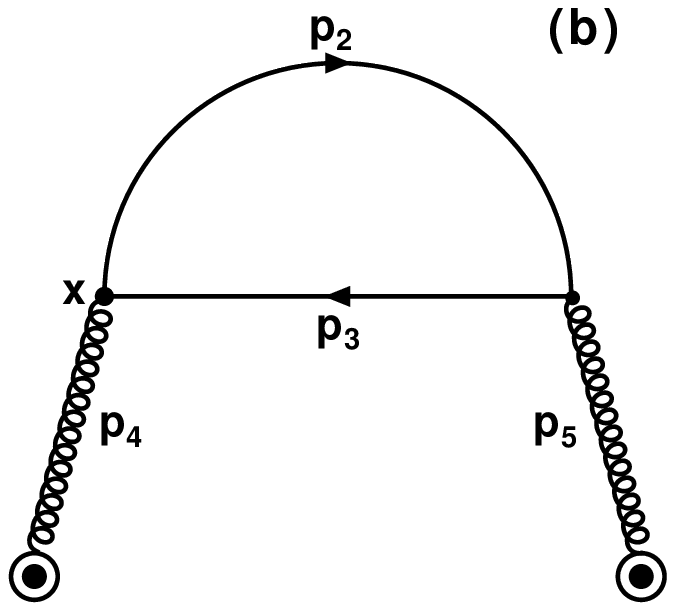}
}
\caption{The Feynman diagrams contributing to $\langle \partial_\mu T^{\mu\nu}\rangle$ in the presence of the interaction term, $A_\mu^a \, j^\mu_a$.  The contribution from the energy-momentum tensor and external current insertions are discussed in the text (see also Figure \ref{feyn2}).  The diagram in Figure 1(a) can be traced back to terms in the energy-momentum tensor (see equation (\ref{qemt})) which go as $\bar{\psi}\gamma\partial\psi$, whereas the diagram in Figure 1(b) originates from terms that go as $g\,\bar{\psi}\gamma A\psi$.  
}
\label{feyn1}
\end{figure*}

As mentioned in the Introduction, one must specify how the fast parton couples to the medium in order to investigate the medium response.  The choice that is adopted here is to model the fast parton as an external color current which couples to the Lagrangian:
\eqf{\label{couplesource}
\cur{L} \rightarrow \cur{L} - A_\mu^a \, j^\mu_a,
}
where for the moment I do not specify the explicit form of $j$ ($j^\nu$ should not be confused with the source term, $J^\nu$).  Non-Abelian gauge symmetry is preserved by the replacement made in (\ref{couplesource}) if $D^{ab}_\mu j^\mu_b = 0$.  

In the presence of the interaction term, $A_\mu^a \, j^\mu_a$, the lowest order Feynman diagrams needed to calculate the thermal expectation $\langle \partial_\mu T^{\mu\nu}\rangle$ are shown in Figure \ref{feyn1}.  The diagram in Figure 1(a) arises from terms in (\ref{qemt}) which go as $\bar{\psi}\gamma\partial\psi$, whereas Figure 1(b) arises from terms that go as  $g\,\bar{\psi}\gamma A\psi$.  Note that a two gluon exchange is necessary to couple to the EMT, which is a color singlet quantity (a two photon exchange is also necessary in QED, from Furry's theorem).  The convention used here is that the standard Feynman quark-gluon vertex contributes $i g \gamma^\mu t^a$ and the Feynman gauge is used for gluon propagators.

\begin{figure}
\centerline{
\includegraphics[width = 0.35\linewidth]{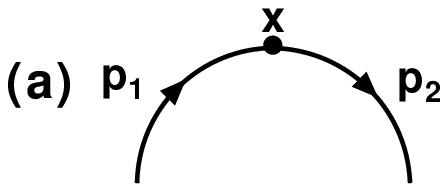}\hskip0.05\linewidth
\includegraphics[width = 0.26\linewidth]{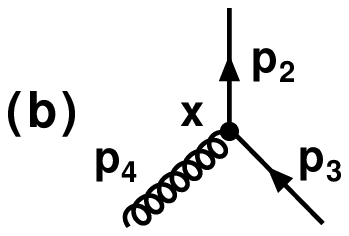}\hskip0.05\linewidth
\includegraphics[width = 0.17\linewidth]{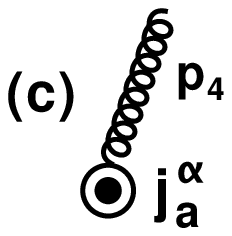}
}
\caption{Contributions to the Feynman diagrams of Figure \ref{feyn1} which come from inserting the energy-momentum tensor and external current interaction term.  Their values and how they are obtained are discussed in the text, see specifically equations (\ref{fig2a}) - (\ref{sourcerule}).}
\label{feyn2}
\end{figure}

One must also determine what the correct Feynman rules for the EMT and external current are.  To determine these non-standard contributions, which are isolated in Figure \ref{feyn2}, one first assigns the appropriate momentum to each derivative.  Recalling that one is here interested in $\langle \partial_\mu T^{\mu\nu}\rangle$ and that the final result is Fourier transformed into position space, it is found that the value of Figure \ref{feyn2}(a) is \cite{Neufeld:2010xi}
\eqf{\label{fig2a}
&\frac{i e^{-i x \cdot(p_1 - p_2)}}{4} \left((p_2^2 - p_1^2) \gamma^\nu  + p_1^\nu (3 \slashed{p}_2 + \slashed{p}_1) -  p_2^\nu(3\slashed{p}_1 + \slashed{p}_2) \right).
}
One applies the same procedure to Figure \ref{feyn2}(b) where I choose the convention that the gluon momentum flows away from the external current (or into $x$ and any vertex) in all cases.  The result is \cite{Neufeld:2010xi}
\eqf{\label{fig2b}
-&i g \,e^{-ix\cdot(p_4 + p_3 - p_2)}\,(p_4 + p_3 - p_2)_\mu \frac{\left(\gamma^\nu j_a^\mu + \gamma^\mu j_a^\nu - 2 g^{\mu\nu}\slashed{j_a}\right)t^a}{2}
}
which is only valid when the gluon field in Figure \ref{feyn2}(b) connects with the source.

Finally, for the external color current in Figure \ref{feyn2}(c), one has very generally
\eqf{\label{sourcerule}
-i\int d^4 z \,j^\alpha_a(z) \, e^{i z\cdot p_4}
}
for the case of an external current which contains one power of $g$.  As mentioned in the Introduction, the focus here will be on an asymptotically propagating fast parton.  In that case $j^\mu_a = g Q^a(t) U^\mu\,\delta^3(\bald{z} - \bald{u}\,t)$ where $\bald{u}$ is the fast parton's velocity and $U^\mu = (1,\bald{u})$.  $g Q^a(t)$ is the charge of a classical particle in QCD (defined by $Q^a_i\,Q^a_j = \delta_{ij} C_{2i}$, with $C_{2i}$ the quadratic Casimir in representation $i$ (3 for a gluon, 4/3 for a quark)) and $Q^a(t)$ evolves in time according to Wong's equations \cite{Wong:1970fu}.  The time dependence of $Q^a (t)$ is of higher order in $g$ and will not be considered here.  For this choice of $j^\mu_a$ Figure \ref{feyn2}(b) simplifies to
\eqf{\label{easysourcerule}
-2\pi i \,g \, Q^a U^\alpha\,\delta(p_4\cdot U).
}

As discussed in the Introduction, the approach used in this paper can easily be extended to consider more realistic scenarios than the single asymptotic parton considered here.   One can modify the interaction term to consider back to back fast partons created in an initial hard interaction or incorporate the medium induced evolution of a primary fast parton into a full parton shower through radiative processes.

To obtain the Green's function structure of Figure \ref{feyn1} I use the conventions of real time thermal field theory as outlined by Das \cite{Das:1997gg}.  The details of the calculation will not be presented in this proceedings, but will be presented in a follow-up publication in which medium gluons will also be considered.  Using standard Feynman rules for real time thermal field theory in addition to the special rules presented in equations (\ref{fig2a}) - (\ref{sourcerule}) the result for the source term is obtained as \cite{Neufeld:2010xi}
\eqf{\label{fullsourceterm}
\langle \partial_\mu T^{\mu\nu}(x)\rangle &= -4 i \,N_F\,C_2\,g^4 \int \frac{d^4 p_3\,d^4 p_4\,d^4 p_5}{(2\pi)^{9}} e^{-i x \cdot(p_4 + p_5)} \delta(p_3^2)n_F(p_3)G_R(p_4)G_R(p_3 + p_4)G_R(p_5) \\
&\times\delta(p_4\cdot U)\delta(p_5\cdot U) \left[2\,(p_3\cdot U)^2\,p_5^\nu - U^2 \,p_3\cdot p_4\,p_5^\nu - U^\nu\,(p_3\cdot U)(2 p_3\cdot p_5 + p_4\cdot p_5)\right]
}
where $G_{R}(p) = (p^2 + i\epsilon p^0)^{-1}$ is the retarded Green's function, $n_F(p) = (e^{|p^0|/T} + 1)^{-1}$ is the Fermi distribution function, $T$ is the temperature, and $N_F$ is the number of active flavors.

Equation (\ref{fullsourceterm}) is the central result of this proceedings.  One can obtain an intuitive picture of the underlying physics it contains by considering the Green's function structure.  Reading from left to right in the first line it is seen that a particle from the heat bath (represented by the thermal distribution $ \delta(p_3^2)n_F(p_3)$) absorbs a gluon from the external current (represented by $G_R(p_4)$) and then continues to propagate (shown by $G_R(p_3 + p_4)$) until it absorbs another gluon from the external current (represented by $G_R(p_5)$).  A final propagator representing the reabsorption of the particle by the heat bath (which would be given by $G_R(p_3 + p_4 + p_5)$) has been cancelled by the momentum structure of $\langle \partial_\mu T^{\mu\nu}\rangle$.

It is useful to consider certain approximations and limits of equation (\ref{fullsourceterm}).   A kinetic theory calculation of the source term generated by an asymptotically propagating fast parton was performed in \cite{Neufeld:2008hs}.  Within the hard thermal loop (HTL) approximation equation (\ref{fullsourceterm}) should reduce to the kinetic theory result (the HTL approximation is formally equivalent to the Vlasov equation \cite{Kelly:1994dh}).  The interested reader can verify (as this author has done) that (\ref{fullsourceterm}) indeed reproduces the result of \cite{Neufeld:2008hs}.  This verification gives confidence to the approach introduced in this proceedings.

I now consider an application of (\ref{fullsourceterm}) by considering the source term ansatz discussed in \cite{neufrenk}.  The authors of \cite{neufrenk} proposed the following simple form:
\eqf{\label{simplesource}
\langle \partial_\mu T^{\mu\nu}(x)\rangle \approx  \frac{d E}{d t}\left(U^\nu - \lambda \partial^\nu \right)\delta(\bald{x} - \bald{u} t).
}
The utility of (\ref{simplesource}) is that one can encode basic features of the source term in a compact way.  The dimensionfull coefficient $\lambda$ parameterizes local contributions from the source, that is, terms which globally integrate to zero (and for instance do not contribute to $dE/dt$) but may still be important for exciting the medium.  It was found in \cite{neufrenk} that within linearized hydrodynamics a double peaked structure in the azimuthal emission spectrum associated with the source given by (\ref{simplesource}) only appeared for rather large values of $\lambda$ (on the order of 0.5 fm or higher for 20 GeV total energy deposited into the medium).  In what follows, I will extract $\lambda$ from equation (\ref{fullsourceterm}).

To extract $\lambda$, one needs the relation \cite{Neufeld:2010xi}
\eqf{\label{lamform}
\lambda = \frac{-1}{\frac{dE}{dt}}\int d^3 x\,x\,\langle \partial_\mu T^{\mu x}(x)\rangle
}
where I have chosen to use the $x$ component of the source to obtain $\lambda$ (according to the ansatz of (\ref{simplesource}) any component would work).  $dE/dt$ here is the collisional energy loss rate, which can be obtained from the source term as
\eqf{\label{energyconservation}
\frac{d E}{d t} = \int d^3 x \, \langle \partial_\mu T^{\mu 0}(x)\rangle.
}
For this proceedings I will use the HTL limit for the collisional energy loss:
\eqf{\label{dedthtl}
\frac{d E}{d t}_{\text{HTL}} &= \frac{m_D^2\,C_2\,\alpha_s}{2}\left(1 - \frac{\tanh^{-1} [u]}{\gamma^2 \,u}\right)\ln\frac{p_{max}}{p_{min}}
}
where $\gamma = (1-u^2)^{-1/2}$ and based on the HTL approximation and physical reasoning $p_{max} \sim T$ and $p_{min}\sim m_D$.  For simplicity, I will use the relativistic limit ($\gamma\gg1$) of (\ref{dedthtl}).

With $dE/dt$ in hand, $\lambda$ can be obtained directly from (\ref{fullsourceterm}) by using (\ref{lamform}) and an integration by parts of the general form:
\eqf{\label{partsgrad}
\int d^3 x \,d^3 p\,x\, e^{i \bald{p} \cdot \bald{x}}\,f(\bald{p}) = i \int d^3 x \,d^3 p\,e^{i \bald{p} \cdot \bald{x}}\,\partial_{p_x} f(\bald{p}).
}
Care must be taken in using (\ref{partsgrad}) because $\lambda$ has an infrared divergence.  The simplest way around this, and what I will adopt here, is to introduce a mass term in the Green's functions $G_R(p_4)G_R(p_5)$ in (\ref{fullsourceterm}) given by the thermal gluon mass $m^2 = m_D^2/2$.

It is also necessary to consider carefully the momentum scales involved in the integration.  For consistency, one should only apply the full result of equation (\ref{fullsourceterm}) down to some momentum scale $|p_4| \sim q^*$ where $g T \ll q^* \ll T$.  When $|p_4| \sim g T$ the full result contains contributions from higher order terms and only the HTL approximation is consistent.  From a practical point of view in the calculation of $\lambda$ this means dividing terms in the integration such that
\eqf{
\lambda &= \lambda_\text{F} \text{ when } |p_4|\geq q^*, \\
\lambda &= \lambda_{\text{HTL}} \text{ when } |p_4|\leq q^*,
}
where $ \lambda_\text{F}$ is obtained from (\ref{fullsourceterm}) and $\lambda_{\text{HTL}}$ is from taking the HTL approximation of (\ref{fullsourceterm}).  The final result should be independent of $q^*$ (this type of analysis for the separation of scales in calculations of finite temperature field theory was introduced in \cite{Braaten:1991dd}).

\begin{figure}
\centerline{
\includegraphics[width = 0.55\linewidth]{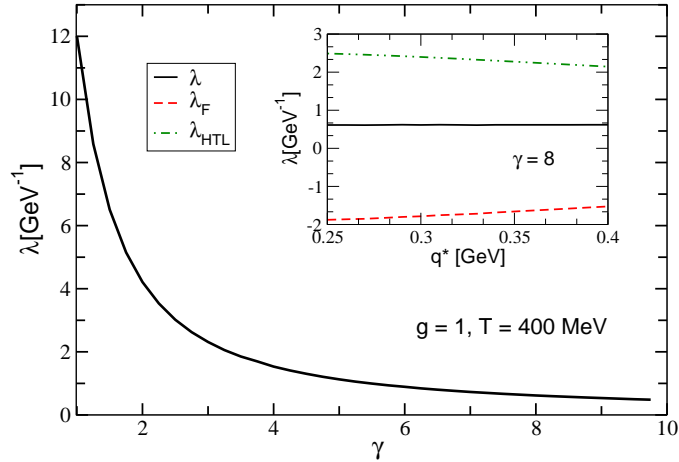}
}
\caption{In the source term of equation (\ref{simplesource}) $\lambda$ couples directly to sound modes and is important for generating observable Mach cone signals.  The contribution to $\lambda$ from the source term (\ref{fullsourceterm}) is plotted above for $g = 1$ and $T = 400$ MeV \cite{Neufeld:2010xi}.  It is clear that $\lambda$ drops as a function of $\gamma = E/m$ which could have implications for the trigger $p_T$ dependence of azimuthal dihadron correlation measurements (see discussion in text).   The inset shows that the result for $\lambda$ is largely independent of the separation parameter, $q^*$ (see text for details).
}
\label{lambda}
\end{figure}

With these technicalities in place, I will present the result for $\lambda$ and further show that it is independent of $q^*$.  The result is presented in Figure \ref{lambda} as a function of $\gamma = (1-u^2)^{-1/2}$ for $g = 1$ and $T = 400$ MeV.  It is interesting that $\lambda$ falls rather sharply as a function of $\gamma$.  As was mentioned above, $\lambda$ parameterizes local contributions from the source and is important for exciting the medium.  In particular, $\lambda$ couples directly to sound modes (and not to diffusive modes) when (\ref{simplesource}) is used as a source term for hydrodynamics.  It was found that large values of $\lambda$ were crucial to the appearance a conical Mach-like emission spectrum in \cite{neufrenk}.  The experimental implication of the dependence of $\lambda$ on $\gamma$ as shown in Figure \ref{lambda} could be found in the trigger $p_T$ dependence of measurements of azimuthal dihadron particle correlations.  In particular, a conical emission pattern would be less likely to be observed for increasing trigger $p_T$, which indeed seems to be the case \cite{Adare:2010ry}.  The inset of Figure \ref{lambda} shows that the result is largely independent of the separation parameter, $q^*$, introduced above.

\section{Conclusion}

In this proceedings I have presented a new approach for coupling the hard physics of fast parton propagation to the underlying thermal medium.  The principle idea is to couple the fast parton to the QGP via an interaction term in the Lagrangian.  The dynamics of the medium in the presence of the fast parton are then directly obtained by taking the thermal expectation value of the energy-momentum tensor.   

Using this approach the source term for a medium of thermalized quarks in the presence of an asymptotically propagating fast parton was obtained and presented in equation (\ref{fullsourceterm}).  It was then found that local excitations, which may be important for generating observable Mach cone signatures, fall sharply as a function of $\gamma$, as shown in Figure \ref{lambda}.  This may have implications for the trigger $p_T$ dependence of measurements of azimuthal dihadron particle correlations in heavy-ion collisions.  In particular, a conical emission pattern would be less likely to be observed for increasing trigger $p_T$, which may indeed be the case \cite{Adare:2010ry}.  

A future publication will include medium gluons and present the details of the calculation \cite{neuvitev}.  The approach introduced here is general and can be modified to consider the medium response to tagged jets \cite{lhcrad} or medium induced full parton showering constrained by realistic radiative energy loss calculations \cite{grig}.

\section{Acknowledgments}
I wish to thank the organizers and participants of the 2011 Winter Workshop on Nuclear Dynamics for a great conference.

\end{document}